\documentclass[a4paper,12pt]{article}
\usepackage{Stylefile}
\doublespace 

\begin{document}


\title{\vspace* {2.0cm} \ProgName: A Program for Batch Processing and \\
Visualization of Powder Diffraction Data \vspace{2.0cm} }

\author{Taha Sochi\footnote{
ScienceWare, PO Box 293, London, N1 5UY, UK. Email: \texttt{}{\color{blue}
tahasochi@scienceware.net}.} \vspace{6cm}}

\date{2009}

\maketitle

\thispagestyle{empty} %


\pagenumbering{roman}

\newpage
\phantomsection \addcontentsline{toc}{section}{Contents} %
\tableofcontents %
%

\phantomsection \addcontentsline{toc}{section}{List of Figures} %
\listoffigures

\phantomsection \addcontentsline{toc}{section}{List of Tables} %
\listoftables

\pagestyle{headings} %
\addtolength{\headheight}{+1.6pt}
\lhead[{Chapter \thechapter \thepage}]%
{{\bfseries\rightmark}}
\rhead[{\bfseries\leftmark}]%
{{\bfseries\thepage}} 
\headsep = 1.0cm               %

\pagenumbering{arabic}

\newpage
\section{Abstract} \label{Abstract}
In this article we report the release of a new program for batch processing and visualization of
powder diffraction data. The program, which is free-of-charge for non-commercial use and can be
obtained with its detailed documentation from our website \url{www.scienceware.net}, is currently
in use by a number of researchers in University College London, University of Manchester, Utrecht
University in the Netherland, European Synchrotron Radiation Facility (ESRF), and Diamond Light
Source. The software is designed for the treatment of large volume of powder diffraction data,
especially those obtained from the new generation of synchrotron detectors. The program has a great
potential for future development to be a workbench for powder diffraction work.

\newpage
\section{Introduction} \label{Introduction}
The principal objective of `\ProgName' project is to develop a computer code for batch refining and
bulk analysis of large volume of diffraction data sets, mainly those obtained from synchrotron
radiation sources. Such a facility greatly assists studies on various materials systems and enables
far larger detailed data sets to be rapidly interrogated and refined. Modern detectors coupled with
the high intensity X-rays available at synchrotrons have led to the situation where data sets can
be collected in ever shorter time scales and in ever larger numbers. Such large volume data sets
pose a data processing problem which is set to augment with the current and future instrument
development. So far, `\ProgName' has achieved a number of its original objectives. In a recent
assessment by the author of the code (i.e. the author of this article) it was concluded that the
time has come for the program to be released for the use by the wider scientific community, in
particular powder diffractionists and synchrotron users.

\newpage
\section{\ProgName} \label{ProgName}
\ProgName\ is a high throughput software to manage, process, analyse and visualise powder
diffraction X-ray data, especially those obtained from synchrotron facilities such as Diamond
\cite{Diamond} and ESRF  \cite{ESRF}. The name `\ProgName' comes from the original name `EasyEDD'
which was adopted for historical reasons as the software was developed initially for the users of
station 16.4 of Daresbury Synchrotron Radiation Facility (SRS) \cite{SRS}, which was closed in
August 2008. As the program has eventually evolved to be more general and can be used for
processing \ADDl\ (ADD) data as well as \EDDl\ (EDD) data, the name `\ProgName' was adopted to
consider this extension. In fact there is no restriction on the program to be used for general
applications not related to synchrotron and powder diffraction, as the program is sufficiently
general for the use of processing any data having the correct format. One of these formats is a
generic $x$-$y$ entries which can be used for any data type. The program is written in C++
programming language and uses a hybrid approach of procedural and object oriented programming
methodologies. Its main purpose is to process large quantities of data files with ease and comfort
using limited time and computing resources. \ProgName\ combines Graphic User Interface (GUI)
technology (e.g. wizards, dialogs, tooltips, color coding, context menus, and so forth) with
standard scientific computing techniques. The ultimate objective for \ProgName\ is to become a
workbench for batch processing and analysis of scientific data, especially from synchrotron and
powder diffraction applications. Currently, five input data file formats are supported. The code
can be easily extended to support other data formats. The current supported formats are:

\begin{enumerate}

\item Generic $x$-$y$ format: this is a simple $x$-$y$ format where the first column of the data file contains the $x$
values (e.g. energy or channel number or scattering angle) while the second column contains the $y$
values (usually count rate). The rest of each line (which can contain other data such as count
error) is ignored. This format is very general and can be used for reading, mapping and analysing a
variety of data files.

\item Diamond MCA format: this format is used by some detectors in the Diamond Light Source of
Rutherford Appleton Laboratory (RAL) where the file contains the $y$ values (count rate) only as a
function of an implicit channel number with possible redundant header and footer lines.

\item ESRF MCA format: this is one of the formats that are in use by the ESRF detectors. The files have
only headers to be ignored, and each 16 data entries are on a single text line ending with
backslash `$\backslash$'. Again, the data in the ESRF MCA files are intensity versus implicit
channel number.

\item ERD detector format: this is the format of the 2D Energy Resolving Detector (ERD). Each row in the
ERD files corresponds to an event taking place (a photon being detected by a specific pixel). The
main data in each row are the values in volt corresponding to the energy of events and the pixel
number in which the event took place. The pixels are arranged in a 2D array and hence a map file
that depicts the physical layout of the detector channels is required. The map file contains the
number of rows, the number of columns and the voltage bin size. This is followed by the channels
index map which mimics the physical layout of the 2D detector. All the lines with the same pixel
number are grouped together and the energies for each pixel are processed as spectra. The
$x$-values of these spectra are the bin number while the $y$-values are the count rate (number of
events in a specific bin). Because the energy is not quantised, the binning process is applied
before creating a spectrum. The bin size is determined by the user, and is input through the map
file. The routine that implements this operation is general with regard to the detector dimension,
so it can be used for a hypothetical 100$\times$110 detector.

\item SRS 16.4 format: this is the format of the data files obtained from station 16.4
of Daresbury Synchrotron Radiation Source in EDD mode. These files are classified as scalars and
vectors, where the scalars contain data about the diffraction measurements while the vector files
contain the actual measurements. Each vector file contains a number of EDD spectra ($\sim20$)
stored as intensity versus implicit channel numbers. The number of channels depends on the type of
Multi-Channel Analyser (MCA) detector in use (normally 4000). Also each set of measurements
consists of a single scalar file and three vector files, one from each of the three TEDDI
detectors.

\end{enumerate}

One of the main functionalities of \ProgName\ in its current state is to read and map data files.
What is required from the user is to deposit the data files of a particular format in a directory
and invoke the relevant read data function. In the case of SRS files where the data files have a
highly structured format, the data files are read and automatically recognised (e.g. SRS, scalars
or vectors), and therefore non-SRS data files in the source directory will have no effect as they
will be recognised and ignored. For the other four formats, where the structure is relatively
generic and cannot be automatically recognised by the program, the program relies on the user for
identifying the files. The program therefore assumes that all the files in the source directory are
valid data files of the selected format.

On reading the data files, the data are stored in memory and mapped on a 2D color-coded tab, as
seen in Figure (\ref{MainWindow2}). Multiple tabs from different data sources can be created at the
same time. The tabs can also be removed collectively or individually in any order. In the following
sections we outline the main components of \ProgName.

\newpage
\subsection{Main Window} \label{MWin}
\begin{figure} [!h]
\includegraphics
[scale=0.9] {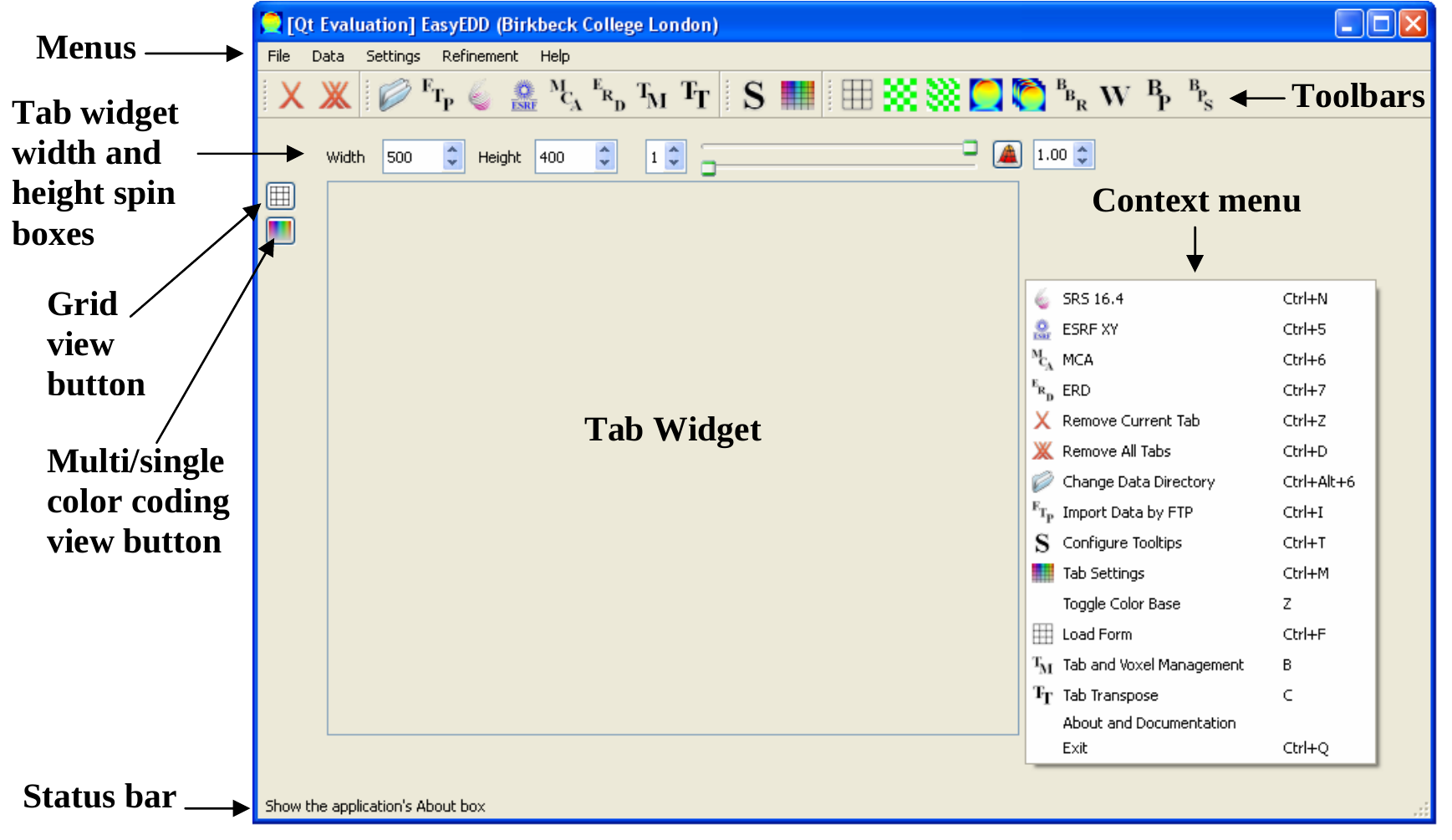}
\caption[\ProgName\ main window] {\ProgName\ main window.} \label{MainWindow}
\end{figure}

This is a standard GUI window with menus, toolbars, a status bar, context menus and so on. Figure
(\ref{MainWindow}) displays a screenshot of the main window. The basic functionality of the main
window is to serve as a platform for accessing and managing the other components with their
specific functionalities.

The main window has several menus which are the main access tool to the program utilities. Some of
these menus also have submenus. The main window menus contain all the principal items of the
program with their icons and shortcuts. The data in any tab can also be exported in a generic text
format for the use in other applications, such as Excel, or for archiving especially after
introducing corrections and modifications. Intensities, refined parameters and statistical
residuals after curve-fitting can also be exported separately to generic text files for further
processing or for visualisation by other applications such as MATLAB. The written data are
structured in a 2D matrix with the same dimensions as the tab.

The spectrum of each voxel, as represented by a cell in a tab, can be plotted using a 2D plotter.
The sum of all patterns in a tab can also be plotted by the 2D plotter to compare with the
individual patterns and to charcterise the overall behaviour. Rows, columns and cells in each tab
can be managed and manipulated by 12 functions which include delete, copy, exchange, rotate
clockwise and anticlockwise and so on. Moreover, the tab can be transposed by reflection across the
$y=-x$ line.

\subsection{Tab Widget} \label{MWTW}
\begin{figure} [!h]
\centering
\includegraphics
[scale=0.9] {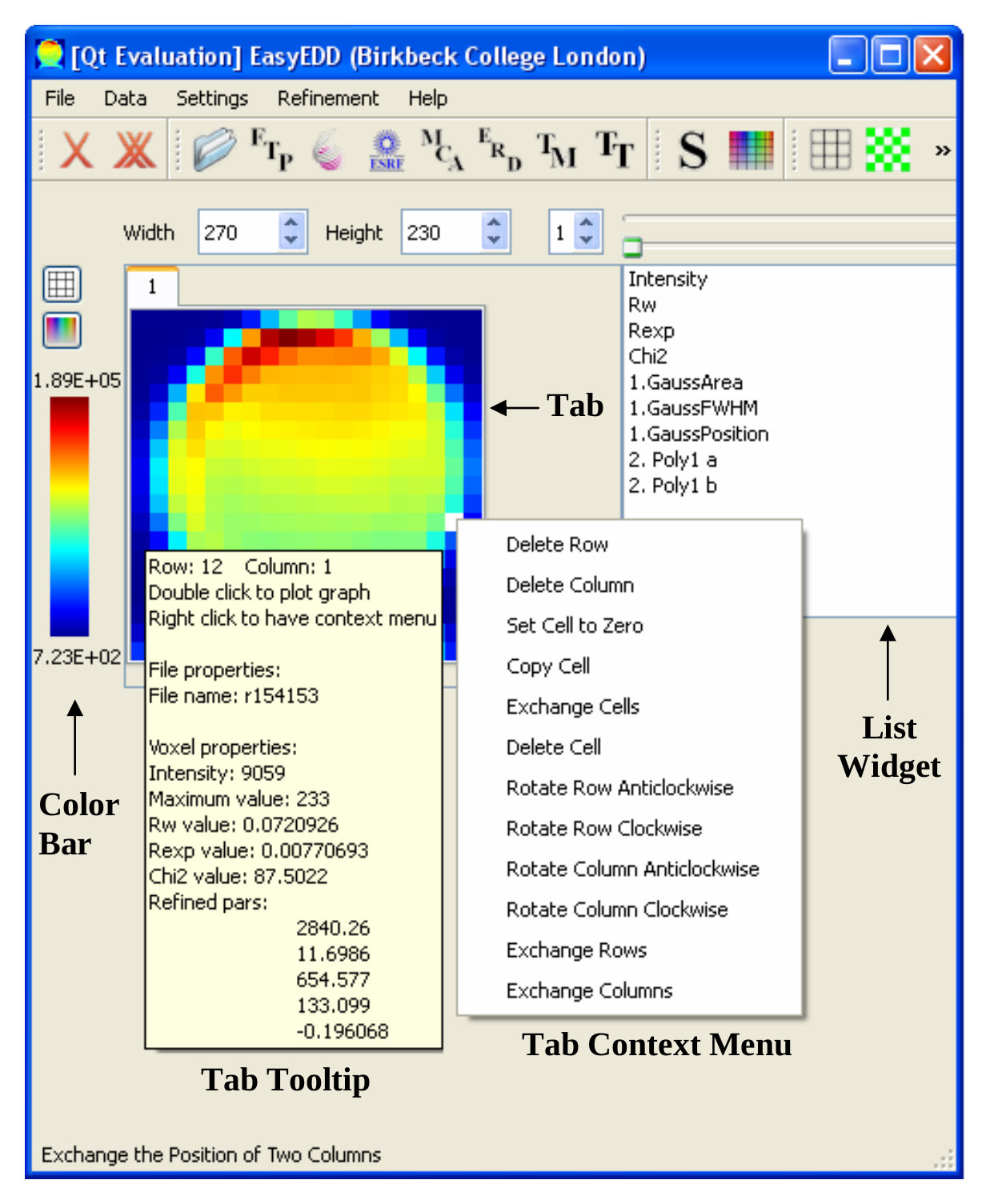}
\caption[\ProgName\ main window with tab tooltip, tab context menu and list widget after reading
data and performing batch fitting] {\ProgName\ main window with tab tooltip, tab context menu and
list widget after reading data and performing batch fitting.} \label{MainWindow2}
\end{figure}

This is a widget (Figures \ref{MainWindow} and \ref{MainWindow2}) that can accommodate a number of
2D color-coded scalable tabs for voxel mapping with graphic and text tooltips to show all essential
file and voxel properties. On double-clicking a cell in a tab in the tab widget the plotter will be
launched, if it was not launched already, and the pattern of the selected cell is plotted. The
plotter is dynamically updated by hovering the mouse cursor over the tab so that the plotter
displays the pattern of the cell where the mouse cursor is currently placed. The tabs have graphic
and text tooltips and a context menu for managing the rows, columns and cells in the tab.

\newpage
\subsection{2D Plotter} \label{2DPlotter}
\begin{figure} [!h]
\centering
\includegraphics
[width=0.9\textwidth] {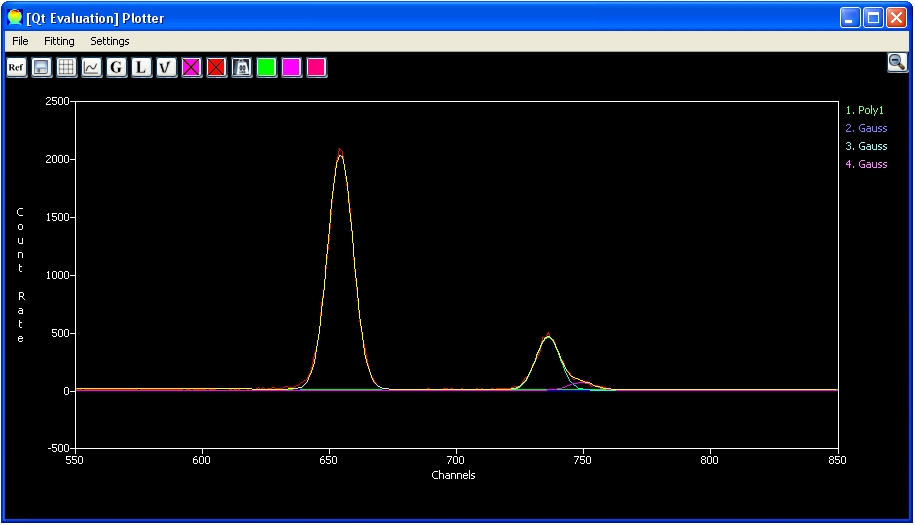}
\caption[Plotter] {The 2D plotter.} \label{plotter}
\end{figure}

This is a 2D plotter to obtain a graph of intensity for any voxel in the current tab by clicking on
its cell. It is also used to create basis functions and forms for curve-fitting. A screenshot of
the plotter is shown in Figure (\ref{plotter}). The 2D plotter capabilities include:

\begin{itemize}

\item Creating and drawing fitting basis functions (polynomials of order $\leq$ 6 that pass through
a number of selected points, Gauss, Lorentz and pseudo-Voigt profiles, as presented in Table
\ref{lineShapeFuncs}) by mouse click or by press and drag actions. The fitting basis functions can
also be modified and removed.

\item Non-linear least squares curve-fitting by Levenberg-Marquardt algorithm.

\item Saving the plotter image in a number of different formats (e.g png, bmp, jpg, jpeg, xpm and pdf).

\end{itemize}


\begin{table} [!h]
\centering %
\caption[Some of the common shape functions used in \ProgName\ to describe individual peak
profiles]
{Some of the common shape functions used in \ProgName\ to describe individual peak profiles. $A$ is
area under peak, $F$ is FWHM, $X$ is position of peak and $m$ is a dimensionless mixing factor ($0
\leq m \leq 1$).}
\label{lineShapeFuncs} %
\vspace{0.5cm} %

\begin{tabular}{ll}
\hline
{\bf Function} \verb|                                | & {\bf Equation} \\
\hline      \vspace{-0.3cm} \\
Gaussian &    $G = \frac{2 \Area}{\FWHM} \, \, \sqrt[]{\frac{{\rm ln}(2)}{\pi}} \, \, e^{^{\frac{-4 {\rm ln}(2)(x-\Pos)^{2}}{\FWHM^{2}}}}$ \\ %
\vspace{-0.3cm} \\
Lorentzian &    $L = \frac{2 \Area /(\FWHM \pi)}{1 + 4(x-\Pos)^{2} / \FWHM^{2}}$ \\ %
\vspace{-0.3cm} \\
Pseudo-Voigt &   $V = \weiFac L + (1-\weiFac) G$ \\ %
\vspace{-0.4cm} \\
\hline
\end{tabular}
\end{table}

\newpage
\subsection{Form} \label{Form}
\begin{figure} [!h]
\centering
\includegraphics
[scale=0.6] {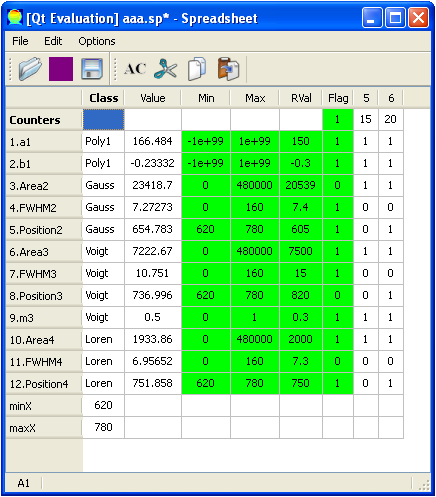}
\caption[Spreadsheet form] {The spreadsheet form.} \label{form}
\end{figure}

The spreadsheet form is mainly used for batch curve-fitting. The idea is that a form is
conveniently prepared within the plotter and saved to the computer hard disc or any other suitable
storage media. It is then imported to the main window for batch fitting a number of cells or tabs
in the tab widget or a number of data sets stored in a directory. The form has two modes of launch,
one from the plotter and the other from the main window. In the first mode the form interacts with
the plotter, that is the form started from the plotter will automatically lists the existing basis
functions in the plotter, while the plotter will be updated on modifying the parameters of the
basis functions in the form. In the second mode the form will serve as a fitting model for batch
curve-fitting. A screenshot of the spreadsheet form is shown in Figure (\ref{form}).

The form consists of a number of menus, toolbars and a status bar, similar to the ones in the main
window. The form also has a number of columns that contain data required for curve-fitting such as
data range; initial fitting parameters; upper and lower limits, boolean flags and values for
applying restrictions on the refined parameters when they exceed acceptable limits; and counters
and boolean flags for controlling the number of iteration cycles and the parameters to be refined
in the least squares fitting process.


\newpage
\subsection{3D Plotter} \label{3DPlotter}
\begin{figure} [!h]
\centering
\includegraphics
[scale=0.4] {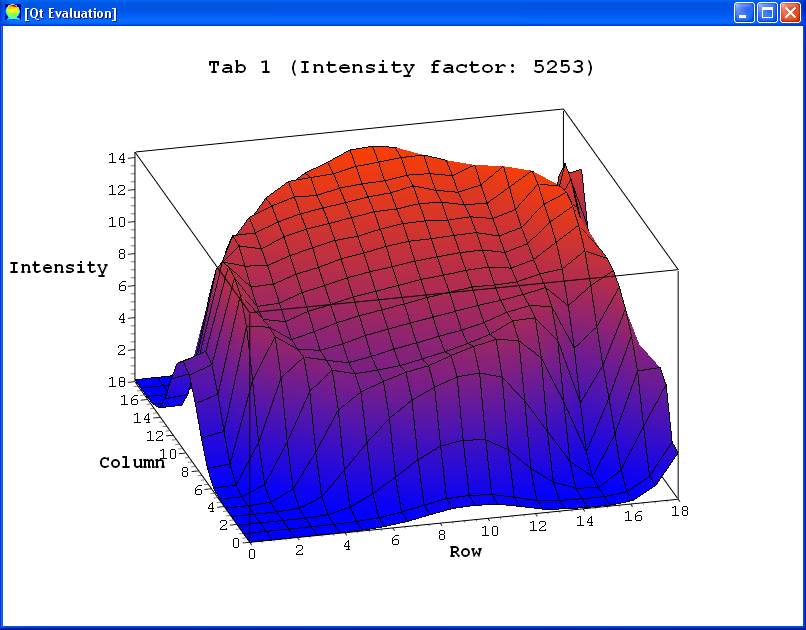}
\caption{The 3D graph of intensity as a function of the voxel position in a tab.} \label{3DGraph}
\end{figure}

The 3D plotter is used to create a 3D graph of the current tab where the $x$ and $y$ axes stand for
the numbers of rows and columns respectively, while the $z$-axis represents the intensity (count
rate) of the voxel, as shown in Figure (\ref{3DGraph}). To keep the proportionality of the plot
dimensions, the intensity is automatically scaled by a scale factor which is displayed beside the
tab number. The 3D plot can be scaled by re-scaling the $z$-axis in the range between 0.01-100. The
plotter responds to the adjustment dynamically. On performing batch refinement in one of its
various modes the residuals and the refined parameters can also be visualised by the 3D plotter.

\newpage
\section{Modules of \ProgName} \label{Modules}
Currently, there are two main modules implemented in \ProgName; these are curve-fitting and back
projection:

\subsection{Curve-Fitting} \label{MWCF}
One of the main modules of \ProgName\ is curve-fitting by least squares using Levenberg-Marquardt
algorithm. Curve-fitting can be performed on a single pattern from the plotter, or as a single
batch process over multiple patterns, or as a multiple batch process over multiple forms and data
sets from the main window. Curve-fitting can be done on a single or multiple peaks using any number
of basis functions with and without polynomial background modeling. The range of the data to be
fitted can also be selected graphically. Some relevant statistical indicators for the fitting
process, as outlined in Table (\ref{statisticalIndic}), are also computed in the curve-fitting
routine. It should be remarked that curve-fitting is a general module that can be used for general
spectral analysis in any application. Hence, \ProgName\ can be used for curve-fitting any data type
from disciplines other than powder diffraction and synchrotron radiation such as astronomical data
or other spectral data. As indicated already, there are three main curve-fitting modes:

\begin{enumerate}

\item Single Curve-Fitting: this is carried out within the plotter on a single pattern, i.e. on the
pattern belonging to the cell that has been plotted on double-clicking the cell. Single
curve-fitting applies to the $x$-range in the current view of the plotter.

\item Batch Curve-Fitting: this operation mode is carried out from the main window. Diffraction
pattern data and a spreadsheet form that contains basis fitting functions should be loaded before
running this operation. Batch curve-fitting applies on the $x$-range of the loaded form.
Curve-fitting in this mode can be performed on a single pattern, a number of randomly selected
patterns from one tab or a number of tabs. It can also be performed on all cells in a tab or a
number of tabs. Batch curve-fitting can also be applied in an overlapping mode by applying
different fitting models on different parts of a tab. The fitting results can then be written to a
text file in an overlapping mode. On running this mode of batch curve-fitting a list widget for
each refined tab is created. These widgets contain items which include intensity, statistical
indicators and refinement parameters. On selecting each one of these items, the current tab color
coding and color bar change to display the variation of the selected item. The 3D plotter will also
display the selected item.

\item Multiple Batch Curve-Fitting: this operation mode is carried out from the main window. It is an
automated multiple batch curve-fitting process that is particularly useful for large-scale
curve-fitting operation. The idea of this routine is to deposit a number of prepared forms and the
folders containing the data sets to be refined in a directory and the program will carry out an
automated batch curve-fitting operation where each data set will be fitted to each individual form
and the fitting data will be saved in a structured directory tree. The residuals and the refined
parameters files, alongside the intensity and other results files for each fitting, will be saved
in a directory bearing the name of the fitting form inside the directory of the particular data
set. A report that contains informative messages about the outcome of the fitting processes will be
written to a file. The use of this operation to replace repeated manual application of batch
curve-fitting process produces more reliable results and can save a huge amount of time.

\end{enumerate}


\begin{table} [!h]
\centering %
\caption[Statistical indicators for powder diffraction pattern refinement] %
{Statistical indicators for powder diffraction pattern refinement. $\yobs$ and $\ycal$ are the
observed and calculated count rate respectively, $\wei$ is the statistical weight, $\NO$, $\NP$ and
$\NC$ are the numbers of observations, adjusted parameters in the calculated model, and applied
constraints respectively, $\Iobs$ and $\Ical$ are the observed and calculated integrated intensity
respectively, and $i$ and $k$ are counting indices.} %
\label{statisticalIndic} %
\vspace{0.5cm} %

\begin{tabular}{ll}
\hline
{\bf Statistical indicator} \verb|                    | & {\bf Definition} \\
\hline      \vspace{-0.3cm} \\
Profile residual &            $\Rp = \frac{\sum_{i} |\yobs_{i} -  \ycal_{i}|}{\sum_{i} \yobs_{i}}$ \\ %
\vspace{-0.3cm} \\
Weighted profile residual &   $\Rwp = \left[ \frac{\sum_{i} \wei_{i}(\yobs_{i} -  \ycal_{i})^{2}}{\sum_{i} \wei_{i}  {\yobs_{i}}^{2}} \right]^{1/2}$ \\ %
\vspace{-0.3cm} \\
Expected residual &           $\Rexp = \left[ \frac{\NO - \NP + \NC}{\sum_{i} \wei_{i}{\yobs_{i}}^{2}} \right]^{1/2}$ \\ %
\vspace{-0.3cm} \\
Bragg residual &              $\RB = \frac{\sum_{k} |\Iobs_{k} -  \Ical_{k}|}{\sum_{k} |\Iobs_{k}|}$ \\ %
\vspace{-0.3cm} \\
Structure factor residual &   $\RS = \frac{\sum_{k} |\sqrt{\Iobs_{k}} -  \sqrt{\Ical_{k}}|}{\sum_{k} |\sqrt{\Iobs_{k}}|}$ \\ %
\vspace{-0.3cm} \\
Goodness-of-fit index         & $\GoF = \left[ \frac{\Rwp}{\Rexp} \right]^{2} =  \frac{\sum_{i} \wei_{i}(\yobs_{i} -  \ycal_{i})^{2}} {(\NO - \NP + \NC)}$ \\ %
\vspace{-0.3cm} \\
\hline
\end{tabular}
\end{table}

\subsection{Back Projection} \label{MWBP}
The purpose of this module is to reconstruct images from sinograms consisting of a set of rotation
versus translation measurements with possible filtering and application of Fast Fourier Transform.
This standard technique for tomographic reconstruction of 2D images from a set of 1D projections
and the angles at which these projections were taken relies on the application of the inverse Radon
transform and the Fourier Slice Theorem. The flexibility and diversity of this operation as
implemented in \ProgName\ (e.g. back projection of intensity or some channels or all channels in a
tab with dynamic display of reconstructed images and possibility of toggling between sinogram and
reconstructed image) make it very useful.

\newpage
\section{Acknowledgement} \label{Acknowledgement}
The author would like to acknowledge valuable contributions from Prof. Paul Barnes and Dr Simon
Jacques. Dr Jacques should also be accredited for being the originator of the main ideas of
\ProgName.

\newpage
\renewcommand{\refname}{}


\newpage
\section{Nomenclature} \label{Nomenclature}
\begin{spacing}{1.5}
\begin{supertabular}{ll}

$\GoF$              &       goodness-of-fit index \\

                    &   \vspace{-0.2cm}\\

$\Area$             &       area under peak \\
$\NC$               &       number of constraints \\
$\FWHM$             &       Full Width at Half Maximum \\
$\Ical$             &       calculated integrated intensity \\
$\Iobs$             &       observed integrated intensity \\
$\weiFac$           &       mixing factor in pseudo-Voigt function \\
$\NO$               &       number of observations \\
$\NP$               &       number of parameters \\
$\RB$               &       Bragg's residual \\
$\Rexp$             &       expected residual \\
$\Rp$               &       profile residual \\
$\RS$               &       structure factor residual \\
$\Rwp$              &       weighted profile residual \\
$\wei$              &       statistical weight \\
$\Pos$              &       position of peak \\
$\ycal$             &       calculated count rate \\
$\yobs$             &       observed count rate \\

                    &   \vspace{-0.2cm} \\

2D                  &       two-dimensional \\
3D                  &       three-dimensional \\
ADD                 &       Angle Dispersive Diffraction \\
EDD                 &       Energy Dispersive Diffraction \\
ERD                 &       Energy Resolving Detector \\
ESRF                &       European Synchrotron Radiation Facility (Grenoble - France) \\
FWHM                &       Full Width at Half Maximum \\
GUI                 &       Graphic User Interface \\
MCA                 &       Multi-Channel Analyser \\
RAL                 &       Rutherford Appleton Laboratory (Didcot - UK) \\
SRS                 &       Synchrotron Radiation Source (Daresbury - UK) \\
TEDDI               &       Tomographic Energy Dispersive Diffraction Imaging \\
\end{supertabular}
\end{spacing}


\section{References} \label{bibliography}

\vspace{-2.1cm}

\begin{thebibliography}{3}

\bibitem[{(2009)}]{Diamond} 
Diamond Light Source website: \url{www.diamond.ac.uk/default.htm}.

\bibitem[{(2009)}]{ESRF} 
European Synchrotron Radiation Facility (ESRF) website: \url{www.esrf.eu/}.

\bibitem[{(2008)}]{SRS} 
Synchrotron Radiation Source (SRS) website: \url{www.srs.ac.uk/srs/}.

\end{thebibliography}
\end{document}